\documentclass[journal]{IEEEtran}

\usepackage[utf8]{inputenc}

\usepackage[pdftex]{graphicx}
\graphicspath{{../imgs/}}
\DeclareGraphicsExtensions{.pdf,.jpeg,.png}

\usepackage{amsmath}
\usepackage{algorithmic}
\usepackage{array}
\usepackage{cite}
\usepackage[caption=false,font=footnotesize]{subfig}
\usepackage{url}
\usepackage{enumitem}
\usepackage[all]{nowidow}

\begin{document}

\title{Speech Dereverberation with Context-aware Recurrent Neural Networks}

\author{João Felipe Santos and Tiago H. Falk}% <-this % stops a space
%\thanks{Insert thanks text here}}


\maketitle

\begin{abstract}
    In this paper, we propose a model to perform speech dereverberation by
    estimating its spectral magnitude from the reverberant counterpart. Our models
    are capable of extracting features that take into account both short and long-term
    dependencies in the signal through a convolutional encoder (which extracts features
    from a short, bounded context of frames) and a recurrent neural
    network for extracting long-term information. Our model outperforms a recently 
    proposed model that uses different context information depending on the reverberation 
    time, without requiring any sort of additional input, yielding improvements of up to
    0.4 on PESQ, 0.3 on STOI, and 1.0 on POLQA relative to reverberant speech.
    We also show our model is able to generalize to real room impulse responses even when
    only trained with simulated room impulse responses, different speakers,
    and high reverberation times. Lastly, listening tests show the proposed method 	         outperforming benchmark models in reduction of perceived reverberation. 
\end{abstract}


\IEEEpeerreviewmaketitle

\section{Introduction}
\IEEEPARstart{R}{everberation} plays an important role in the perceived quality
of a sound signal produced in an enclosed environment. In highly reverberant environments, perceptual artifacts
such as coloration and echoes are added to the direct sound signal, thus drastically reducing speech signal intelligibility, particularly for the hearing impaired \cite{neuman_combined_2010}.  Automatic speech recognition (ASR) performance is also 
severely affected, especially when reverberation is combined with additive noise \cite{yoshioka_making_2012}. To deal with
the distortions in such environments, several types of speech enhancement
systems have been proposed, ranging from single-channel systems based on simple
spectral subtraction \cite{boll_spectral_1979} to multistage systems which leverage
signals from multiple microphones \cite{cauchi_combination_2015}.

Deep neural networks (DNNs) are currently part of many large-scale ASR systems, both as separate acoustic and
language models as well as in end-to-end systems. To make these systems robust to reverberation, different strategies have been explored, such as feature enhancement during a preprocessing stage \cite{mimura_reverberant_2015} and DNN-based beamforming on raw multichannel speech signals for end-to-end solutions
\cite{sainath_factored_2016}.
On the other hand, the application of DNNs to more
general speech enhancement problems such as denoising
\cite{xu_regression_2015}, dereverberation \cite{han_learning_2015}, and source
separation \cite{weninger_discriminatively_2014} is comparatively in early
stages.

Recently, several works have explored deep neural networks for speech enhancement through two main approaches: spectral estimation and spectral masking. In the first, the goal of the neural network is to predict the magnitude spectrum of enhanced speech signal directly, while the latter aims at predicting some form of an ideal mask (either a binary or a ratio mask) to be applied to the distorted input signal.

Most of the published work in the area uses an architecture similar to the one first presented in \cite{han_learning_2015}: a relatively large feedforward neural network with three hidden layers containing several hundred units (1600 in \cite{han_learning_2015}), having as input a context window containing an arbitrary number of frames (11 in \cite{han_learning_2015}) of the log-magnitude spectrum and as target the dry/clean center frame of that window. The output layer uses a sigmoid activation function, which is bounded between 0 and 1, and normalizes targets between 0 and 1 using the minimum and maximum energies in the data. Moreover, an iterative signal reconstruction scheme inspired by \cite{griffin_signal_1984} is used to reduce the effect of using the reverberant phase for reconstruction of the enhanced signal.

In \cite{wu_study_2016}, the authors performed a study on target feature activation and normalization and their impacts on the performance of {DNN} based speech dereverberation systems. The authors compared the target activation/normalization scheme in \cite{han_learning_2015} with a linear (unbounded) activation function and output normalized by its mean and variance. Their experiments showed the latter activation/normalization scheme leads to higher PESQ and frequency-weighted segmental SNR (fwSegSNR) scores than the sigmoid/min-max scheme.

In a follow-up study \cite{wu_reverberation-time-aware_2017}, the same authors proposed a
reverberation-time-aware model for dereverberation that leverages knowledge of
the fullband reverberation time (T60) in two different ways. First, the step size of the short-time Fourier transform (STFT) is
adjusted depending on T60, varying from 2 ms up to 8 ms (the window size is
fixed at 32 ms). Second, the frame context used at the input of the network is
also adjusted, from 1 frame (no context) up to 11 frames (5 future and 5 past
frames). Since the input of the network has a fixed size, the context length is
adjusted by zeroing the unused frames. The model was trained using speech from
the TIMIT dataset convolved with 10 room impulse responses (RIR) generated at a room with fixed geometry
(6 by 4 by 3 meters), with T60 ranging from 0.1 to 1.0 s. The full training
data had about 40 hours of reverberant speech, but the authors also presented
results on a smaller subset with only 4 hours. The model proposed was a feedforward neural network with 3
hidden layers with 2048 hidden units each, trained using all the
different step sizes and frame context configurations in order to find the best
configuration for each T60 value. The authors considered the true T60 value to be
known at test time (oracle T60), as well as estimated using the T60 estimator
proposed by Keshavarz et al. \cite{keshavarz_speech-model_2012}.

% Talk about masking. Emphasize latest Donald Williamson paper, which seems to be the only
% one on dereverb.
The other well-known approach for speech enhancement using deep neural networks is to predict arbitrary ideal masks instead of the magnitude spectrum \cite{wang_training_2014}. The ideal binary mask (IBM) target transforms the speech enhancement problem into a classification problem, where the goal of the model is to predict which time-frequency cells from the input should be masked, and has been shown to improve intelligibility substantially. The ideal binary mask is defined quantitatively based on a local criterion threshold for the signal-to-noise ratio (SNR). Namely, if the SNR of a given time-frequency cell is lower than the threshold, that time-frequency cell is masked (set to zero). Alternatively, the ideal ratio mask is closely related to the frequency-domain Wiener filter with uncorrelated speech and noise. It is a soft masking technique where the mask value corresponds to the local ratio between the signal and the signal-plus-noise energies for each time-frequency cell. Recently, \cite{williamson_complex_2016} has proposed the complex ideal ratio mask, which is applied to the real and imaginary components of the STFT instead of just the magnitude. Models based on mask prediction usually include several different features at the input (such as amplitude modulation spectrograms, RASTA-PLP, MFCC, and gammatone filterbank energies), instead of using just the magnitude spectrum from the STFT representation like the works previously described here. Masks are also often predicted in the gammatone filterbank domain.

In \cite{xiao_speech_2016}, the authors present a model that predicts
log-magnitude spectrum and their delta/delta-delta, then perform enhancement by
solving a least-squares problem with the predicted features, which aims at
improving the smoothness of the enhanced magnitude spectrum. The method was
shown to improve cepstral distance (CD), SNR, and log-likelihood ratio (LLR), but caused slight degradation of the speech-to-reverberation modulation energy ratio (SRMR).
They also reported that the DNN mapping causes distortion for high T60.

Very few studies use architectures other than feed-forward for dereverberation. In \cite{kawahara_speech_2015}, the authors propose an architecture based on long short-term memory (LSTM) for dereverberation. However, they only report mean-squared error and word-error rates for a baseline ASR system (the REVERB Challenge evaluation system) and do not report its effect on objective metrics for speech quality and intelligibility. The method described in \cite{weninger_discriminatively_2014} also uses recurrent neural networks based on LSTMs for speech separation in the mel-filterbank energies domain. Although their system predicts soft masks, similar to the ideal ratio mask, they use a signal-approximation objective instead of predicting arbitrary masks (i.e., the target are the mel-filterbank features of the clean signal, not an arbitrarily-designed mask).

Most current models reported in the literature only explore one of two possible contexts from the reverberant signal. Feed-forward models with a fixed window of an arbitrary number of past and future frames only take into account the local context (e.g. \cite{wu_reverberation-time-aware_2017}) and are unable to represent the long-term structure of the signal. Also, since feed-forward models do not have an internal state that is kept between frames, the model is not aware of the frames it has predicted previously, which can lead to artifacts due to spectral discontinuities. LSTM-based architectures, on the other hand, are able to learn both short- and long-term structure. However, learning either of these structures is not enforced by the training algorithm or the architecture, so one cannot control whether the internal state will represent short-term, long-term context, or both.

In this paper, we propose a novel architecture for speech dereverberation that leverages both short- and long-term context information. First, fixed local context information is generated directly from the input sequence by a convolutional context encoder. We train the network to learn how to use long-term context information by using recurrent layers and training it to enhance entire sentences at once, instead of a single frame at a time. Additionally, we leverage residual connections from the input to hidden layers and between hidden layers. We show that combining short and long-term contexts, as well as including such residual connections, substantially improves the dereverberation performance across four different objective speech quality and intelligibility metrics (PESQ, SRMR, STOI, and POLQA), and also reduces the amount of perceived reverberation according to subjective tests.

\section{Proposed model}
\begin{figure}
    \centering
    \includegraphics[width=0.45\textwidth, height=.5\textheight]{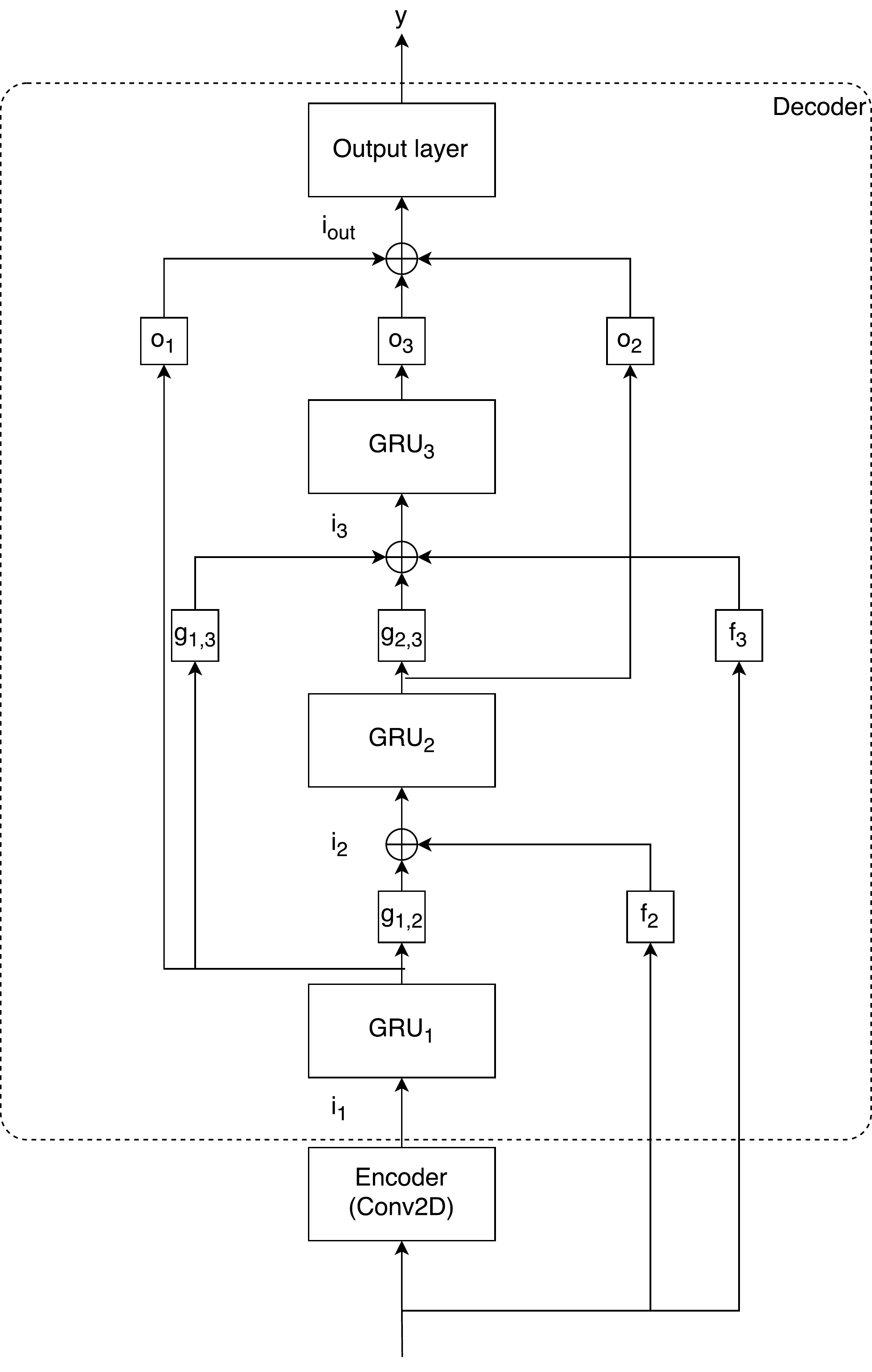}
    \caption{Architecture of the proposed model}
    \label{fig:proposed}
\end{figure}

The architecture of the proposed model can be seen in Fig.~\ref{fig:proposed}. As discussed previously, our model combines both short- and long-term context by using a convolutional context encoder to create a representation of the short-term structure of the signal, and a recurrent decoder which is able to learn long-term structure from that representation. The decoder also benefits from residual connections, which allow each of its recurrent stages to have access both to a representation of the input signal and the state of the previous recurrent layer. Each of the blocks is further detailed in the sections to follow.
\subsection{Context encoder}
As shown in other studies, incorporating past and future frames can help on the
task of estimating the current frame for dereverberation. Most works, however,
use a fixed context window as the input
to a fully-connected layer \cite{han_learning_2015, wu_study_2016}. In this work, we decided to extract local context
features using 2D convolutional layers instead. By using a 2D convolutional
layer, these features encode local context both in the frequency and the time
axis. Our context encoder is composed by a single 2D convolutional layer with
64 filters with kernel sizes of $(21, C)$, where 21 corresponds to the number of frequency bins covered by the kernel and $C$ to the number of frames covered by the kernel in the time axis. In our implementation,
$C$ is always an odd number as we use an equal number of past and future
frames in the context window (e.g., $C = 11$ means the current frame plus 5
past and 5 future frames).  We report the performance of the model for
different values of $C$ in section IV-A. The convolution has a stride of 2 in the frequency
axis and is not strided in the time axis.

\subsection{Decoder}
Following the encoder, we have a stack of three gated recurrent unit (GRU) layers \cite{Cho14} with
256 units each.  The input to the first layer is the output of the convolutional context encoder with all of the channels concatenated to yield an input of shape $(F, T)$, where
\begin{equation*}
    F = 64 \times \left\lfloor\frac{B - N_{context} + 1}{2} +1 \right\rfloor,
\end{equation*}
\noindent 64 is the number of filters in the convolutional layer, $T$ is the number of STFT frames in a given sentence, $B$ is the number of FFT bins (257 in our experiments), and $\lfloor . \rfloor$ is
the \emph{floor} operation (rounds its argument down to an integer).

The outputs to the remaining GRU layers are a combination of affine
projections of the input and the states of the previous GRU layers. Consider
$x_{enc}(t)$ to be the encoded input at timestep $t$, and $h_1(t), h_2(t),
h_3(t)$ as the hidden state at timestep $t$ for the first, second, and third
GRU layers, respectively.  Then, the inputs $i_1(t), i_2(t), i_3(t)$ are as
follows:
\begin{align*}
    i_1(t) &= x_{enc}(t), \\
    i_2(t) &= f_{2}(x(t)) + g_{1,2}(h_1(t)), \\
    i_3(t) &= f_{3}(x(t)) + g_{1,3}(h_1(t)) + g_{2,3}(h_2(t)),
\end{align*}
\noindent where $f_{i}, g_{i,j}$ are affine projections from the input or
previous hidden states. The parameters of those projections are learned during
training, and all projections have an output dimension of 256 in order to match
the input dimension of the GRUs when added together.

Similarly, the output layer following the stacked GRUs has as its input the sum
of affine projections $o_i$ of $h_1, h_2, h_3$:
\begin{align*}
    i_{\text{out}}(t) &= o_1(h_1(t)) + o_2(h_2(t)) + o_3(h_3(t)). %\\
    %y(t) &= W_{\text{out}}i_o(t) + b_{\text{out}}
\end{align*}

\noindent The parameters of those projections are also learned during training.
All of these projections have an output dimension of 256.

\section{Experimental setup}
\subsection{Datasets}
In order to assess the benefits of the proposed architecture for speech
dereverberation, we ran a series of experiments with both the proposed model
and two other models as baselines: the T60-aware model proposed in
\cite{wu_reverberation-time-aware_2017} and a similar model without T60
information that uses a fixed overlap of 16 ms and a fixed context of 11
frames (5 past and 5 future frames). For the T60 aware model, we extracted T60
values directly from the RIRs (oracle T60s) using a method similar to the one used for the ACE Challenge dataset \cite{AceChallenge, karjalainen_estimation_2002}. The
fullband T60 we used was computed as the average of the estimates for the bands
with center frequencies of 400 Hz, 500 Hz, 630 Hz, 800 Hz, 1000 Hz, and 1250
Hz.

For the single speaker experiments, a recording of the IEEE dataset uttered by a
single male speaker was used \cite{loizou_speech_2005}. The dataset consists of 72 lists with 10
sentences each recorded under anechoic and noise-free conditions. We used the
first 67 lists for the training set and the remaining 5 lists for testing.
Reverberant utterances were generated by convolving randomly selected subsets
of the utterances in the training set with 740 RIRs generated using a fast
implementation of the image-source method \cite{lehmann_diffuse_2010}, with T60
ranging from 0.2~s to 2.0~s in 0.05~s steps. Twenty different RIRs (with different
room geometry, source-microphone positioning and absorption characteristics)
were generated for each T60 value. Fifty random utterances from the training
set were convolved with each of these 740 RIRs, resulting in 37,000 files. A
random subset of 5\% of these files was selected as a validation set and used
for model selection and the remaining 35,150 files were used to train the
models. The test set was generated in a similar way, but using a different set 
of 740 simulated RIRs and 5 utterances (randomly selected from the test lists)
were convolved with each RIR.

For the multi-speaker experiments, we performed a similar procedure but using the
TIMIT dataset \cite{timit} instead of the IEEE dataset. The default training
set (without the ``SA" utterances, since these utterances were recorded by all speakers) was used for generating the training and validation sets, and the test set (with the SA utterances removed as well) was
used for generating the test set. The training and test sets had a total of 462 and 168 speakers, respectively. The utterances were convolved with the same
RIRs used for the single speaker experiments. A total of 3696 clean utterances
were used for the training and validation set, and 1336 for the test set. As
with the single speaker dataset, 50 sentences were chosen at random from the training and validation sets (which include all of the utterances from all speakers) and convolved to each of the 740 simulated RIRs to generate the training/validation sets. The test set was generated following the same procedure as used for the single speaker dataset.

Additionally, in order to explore the performance of the proposed and baseline models on realistic settings, we tested the same single speaker models described above with sentences convolved with real RIRs from the ACE Challenge dataset
\cite{AceChallenge}. We used the RIRs corresponding to channels 1 and 5 of the
cruciform microphone array for all of the seven rooms and two microphone positions,
leading to a total of 28 RIRs. The test sentences were the same as for the
experiments with simulated RIRs. The interested reader can refer to \cite{AceChallenge} for more details about the ACE Challenge RIRs.

\subsection{Model Training}
Both the proposed model and the baselines were implemented using the PyTorch library \cite{pytorch} (revision \emph{e1278d4}) and trained using the Adam optimizer \cite{kingma2014adam} with
a learning rate of 0.001, $\beta_1 = 0.9$, and $\beta_2 = 0.999$. The models
were trained for 100 epochs and the parameters corresponding to the epoch with
the lowest validation error were used for evaluation. Since sequences in the
dataset have different lengths, we padded sequences in each minibatch and used masking to compute the MSE loss only for valid timesteps. The models used for evaluation were the ones with the lowest
validation loss amongst the 100 epochs.

\subsection{Objective evaluation metrics}
\label{sec:objmetrics}
We compare the performance of the models using four different objective metrics. The Perceptual Evaluation of Speech Quality (PESQ) \cite{itut_pesq_2001} and Perceptual Objective
Listening Quality Assessment (POLQA) \cite{POLQA_2011} are both ITU-T standards for intrusive speech quality measurement. The PESQ standard was designed for a very limited test scenario (automated assessment of band-limited speech quality by a user of a telephony system), and was superseded by POLQA. The two metrics work in a similar way, by computing and accumulating distortions, and then mapping them into a five-point mean opinion score (MOS) scale. Even though PESQ is not recommended as a metric for enhanced or reverberant speech, several works report PESQ scores for these types of processing and we report it here for the sake of completeness. POLQA is a more complete model and allows measurement in a broader set of conditions (e.g., super-wideband speech, combined additive noise and reverberation). The Short-Time Objective Intelligibility (STOI) \cite{taal2011algorithm} metric is an intrusive speech intelligibility metric based on the correlation of normalized filterbank envelopes in short-time (400 ms) frames of speech. Finally, the speech-to-reverberation modulation energy ratio (SRMR) is a non-intrusive speech quality and intelligibility metric based on the modulation spectrum characteristics of clean and distorted speech, which has been shown to perform well for reverberant and dereverberated speech \cite{Santos_NH_2014,falk_non-intrusive_2010}.

\subsection{Subjective listening tests}
\label{sec:listening}
To further gauge the benefits of the proposed architecture for speech enhancement, we performed a small-scale subjective listening test to assess how effective different methods are in reducing the amount of perceived reverberation. In particular, comparisons with the baseline model Wu2016 were performed as it resulted in improved performance with unmatched speakers relative to the benchmark Wu2017 model. The test protocol we used was the recently proposed MUSHRAR test \cite{mushrar}. 

For the reverberation perception tests, five random samples of the TIMIT test dataset were chosen for each of simulated RIRs with T60s of 0.6, 0.9, 1.2, and 1.5~s and participants were asked to compare the outputs of all models and the reverberant signal to a reference signal. In addition to the model outputs and the reverberant signal, a hidden reference and anchor were used for both experiments: the reference signal was the anechoic signal convolved with an RIR with T60 of 0.2~s and the anchor was the anechoic signal convolved with a RIR with a T60 of 2.0~s. Users were asked to rate the amount of perceived reverberation on a 0-100 scale, with higher values corresponding to higher perceived reverberation. A total of nine participants took part in the test. Tests were performed using a web interface which is freely available online\footnote{Software available at \url{https://github.com/jfsantos/mushra-ruby}}.

\section{Experimental Results}
In this Section, we present the results for four different experiments, namely: 
\begin{enumerate}[label=\Alph*.]
\item Evaluation of the effect of the context size (using the single-speaker, simulated RIR dataset)
\item Comparison between the proposed architecture and baselines on matched conditions (same speaker, simulated RIRs for training and testing)
\item Comparison between the proposed architecture and baselines on mismatched speakers (multispeaker dataset with simulated RIRs)
\item Comparison between the proposed architecture and baselines under realistic reverberation conditions (models trained on single-speaker, simulated RIR dataset, and tested on single-speaker, real RIR dataset).
\end{enumerate}

Since some of the models have a large difference in the number of
learnable parameters, we list the number of parameters for each of the models
that was used in our experiments in Table \ref{tab:nparams}.

\begin{table}
    \centering
    \caption{Number of parameters in each model}
    \label{tab:nparams}
    \begin{tabular}{cc}
        Model & \# of parameters \\
        \hline
        GRU & 4,458,753 \\
        Wu2016 and Wu2017 & 14,711,041 \\
        Proposed without context & 1,838,593 \\
        Proposed, context = 3 frames & 7,429,121 \\
        Proposed, context = 7 frames & 7,434,497 \\
        Proposed, context = 11 frames & 7,439,873
    \end{tabular}
\end{table}

\subsection{Effect of context size}
\begin{figure*}
    \centering
    \subfloat[(a)]{%
        \includegraphics[width=.49\textwidth]{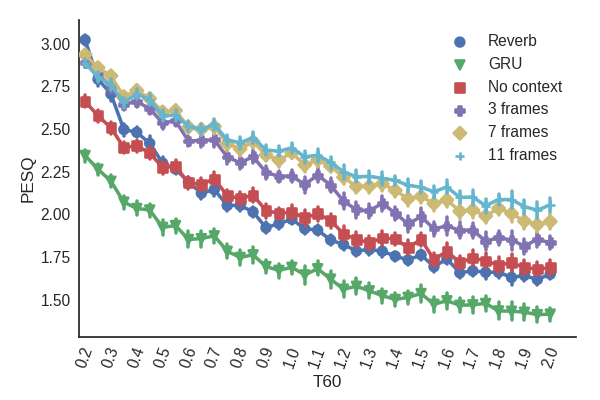}}
        \hfill
    \subfloat[(b)]{%
        \includegraphics[width=.49\textwidth]{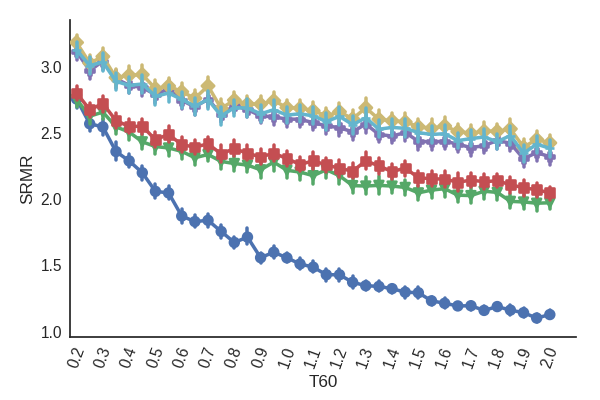}}
        \hfill
    \subfloat[(c)]{
        \includegraphics[width=.49\textwidth]{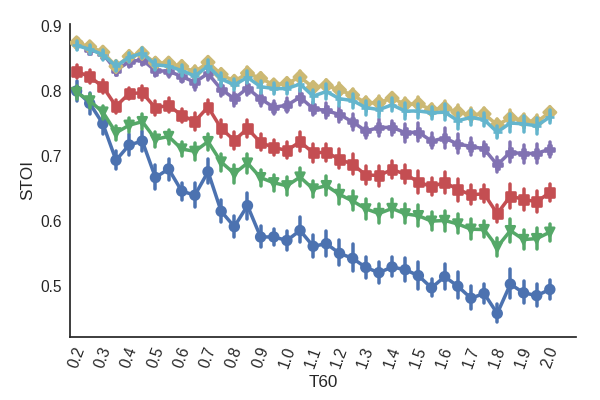}}
    \hfill
    \subfloat[(d)]{%
        \includegraphics[width=.49\textwidth]{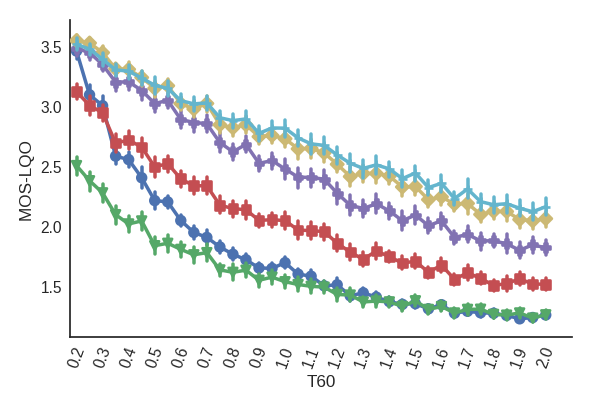}}
    \hfill
    \caption{Effect of context size on scores: (a) PESQ, (b) SRMR,
    (c) STOI, and (d) POLQA.}
    \label{fig:exp2}
\end{figure*}

We first analyze the effect of the context size in the proposed model by
comparing models with context sizes of 3, 7, and 11. We also included a
baseline model based on GRUs without residual connections and a model without
the context encoder and residual connections in these tests. All the models
were trained and tested with single-speaker data convolved with simulated RIRs.
The results can be seen in Figure \ref{fig:exp2}. For this and all of the subsequent experiments in this section, we have subplots for the four objective metrics listed in Section \ref{sec:objmetrics}: (a) PESQ, (b) SRMR, (c) STOI and (d) POLQA in this order. The GRU-only model (without
the context encoder and residual connections) underperforms in all metrics, and
in the cases of PESQ and POLQA, even leads to lower scores than the reverberant
utterances. Adding only the residual connections and no context at all (which
allows the model to perform at real-time) significantly increases the
performance and leads to improvements in all metrics, except for reverberation
times under 400 ms (noticeable in both PESQ and POLQA). Adding the context
encoder, even with a short context of one previous and one future frame (shown
as ``3 frames" in the plots), leads to additional improvements. Adding more
frames (``7 frames" and ``11 frames") improves the performance even further.
However, we see diminishing returns around 7-11 frames, as these two models
have similar performances for SRMR and STOI and only a slight difference in
PESQ and POLQA. Since the model with a context of 11 frames (5 past and 5
future frames) showed the best performance amongst all context sizes, in the
next experiments we show only results with this context size for the proposed
model.

% Comparison of results with different kernel sizes for the convolutional
% kernel (in the time axis)

\subsection{Comparison with baseline models}
\begin{figure*}
    \centering
    \subfloat[(a)]{%
        \includegraphics[width=.49\textwidth]{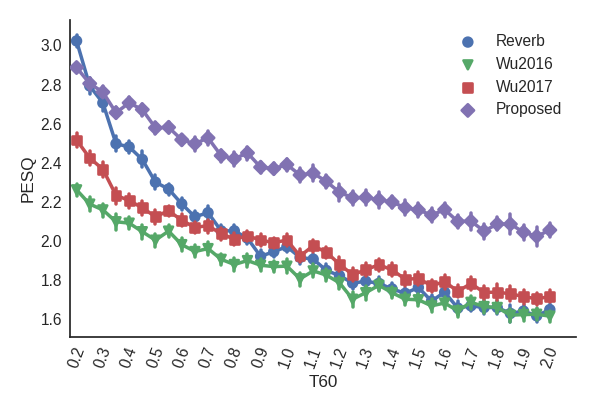}}
    \hfill
    \subfloat[(b)]{%
        \includegraphics[width=.49\textwidth]{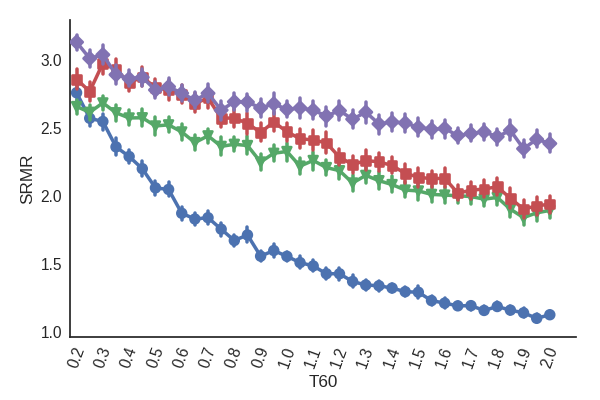}}
    \hfill
    \subfloat[(c)]{
        \includegraphics[width=.49\textwidth]{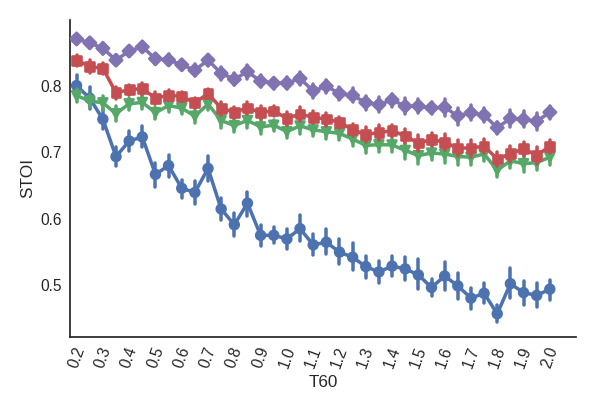}}
    \hfill
    \subfloat[(d)]{%
        \includegraphics[width=.49\textwidth]{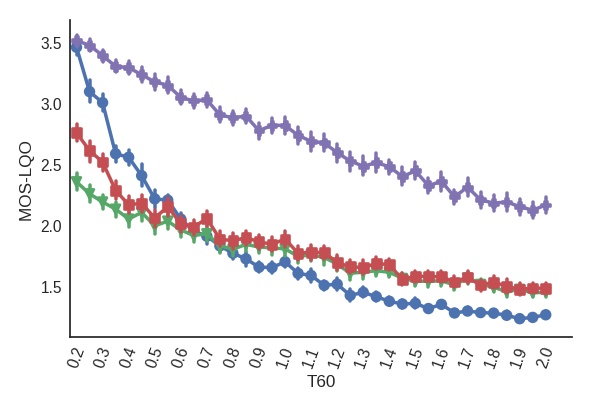}}
    \hfill
    \caption{Single speaker, simulated RIR scores: (a) PESQ, (b) SRMR,
    (c) STOI, and (d) POLQA.}
    \label{fig:exp1}
\end{figure*}

Figure \ref{fig:exp1} shows a comparison between our best model, Wu2016 (fixed
STFT window and hop sizes) and Wu2017 (STFT window and hop sizes dependent on
the oracle T60 values). The interested reader is referred to this manuscript’s supplementary material page to listen to audio samples generated by the proposed and benchmark algorithms \footnote{\url{http://www.seaandsailor.com/demo/index.html}}. A more complete audio demo can be found in \cite{figshare_demo}. It can be seen that our model outperforms these two feed-forward models in
most scenarios and metrics, except for SRMR with $\text{T60} < 0.7$~s, where the results
are very similar. It should be noted, however, that the SRMR metric is less accurate for lower T60s \cite{Santos_NH_2014}.
Even though it uses oracle T60 information, the model Wu2017
does not have a large improvement in metrics when compared to Wu2016,
especially in the STOI and POLQA metrics (which are more sensitive to the
effects of reverberation than PESQ). Using T60 information to adapt the STFT
representation seems to have a stronger effect for lower T60. It should also be
noted that these models lead to a reduction in PESQ scores for lower
reverberation times in the PESQ and POLQA metrics, which is probably due to the
introduction of artifacts. This is also observed for the proposed model, but
only in the PESQ metric and only for $\text{T60} < 0.4$~s. On the other hand, STOI
indicates all models lead to an improvement in intelligibility.

\subsection{Unmatched vs. matched speakers}
\begin{figure*} \centering \subfloat[(a)]{%
    \includegraphics[width=.49\textwidth]{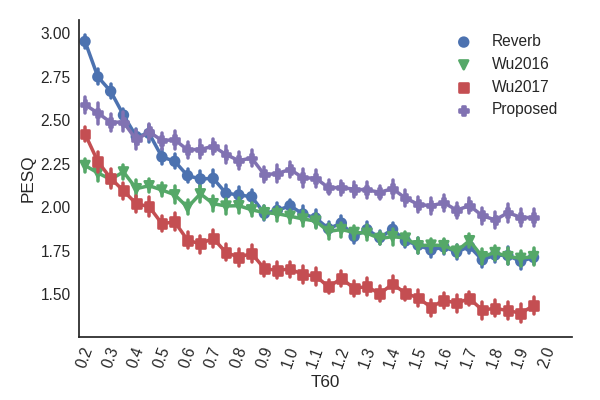}} \hfill
    \subfloat[(b)]{%
        \includegraphics[width=.49\textwidth]{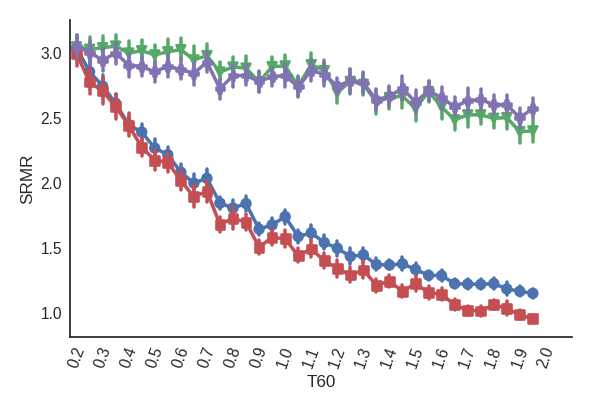}} \hfill
    \subfloat[(c)]{
        \includegraphics[width=.49\textwidth]{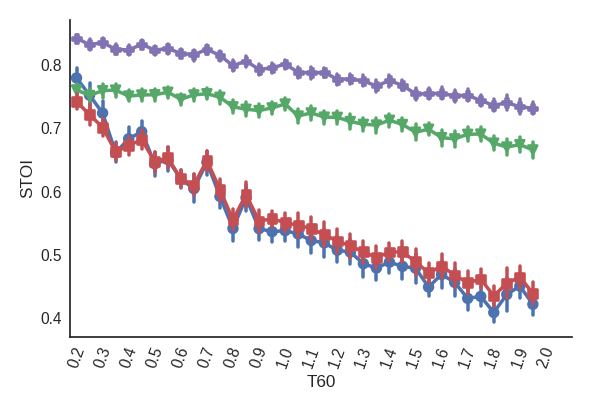}} \hfill
    \subfloat[(d)]{%
        \includegraphics[width=.49\textwidth]{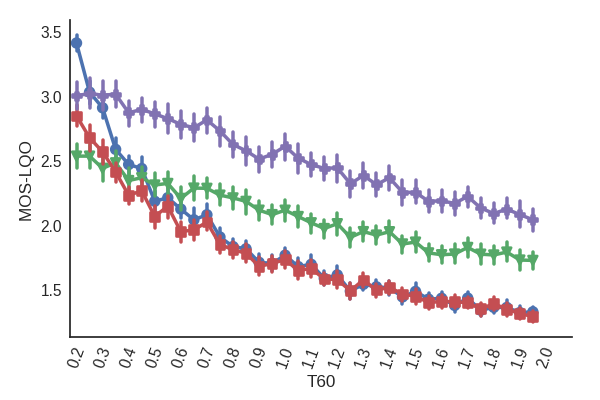}} \hfill
    \caption{Scores for experiment with unmatched speakers:
        (a) PESQ, (b) SRMR, (c) STOI, and (d) POLQA.}
    \label{fig:exp3}
\end{figure*}

Although testing dereverberation models with single speaker datasets allows
assessing basic model functionality, in real-world conditions, training a model
for a single speaker is not practical and has very limited applications. It is
important to evaluate the generalization capabilities of such models with
mismatched speakers, as this is a more likely scenario. Figure \ref{fig:exp3}
shows the results for Wu2016, Wu2017, the proposed model, as well as for the
reverberant files in the test set. Between the baseline models, we can see that
Wu2017 now underperforms Wu2016 in all metrics, and either reduces metric
scores (as for PESQ and POLQA for low T60) or does not change them. However,
the version of the model that does not depend on T60 leads to higher scores.
Although we do not have a clear explanation for this behaviour, we believe the
optimal scores for the STFT hop size and context might depend both on T60 and
speaker, but the Wu2017 model uses fixed values that depend only on T60. Since
the model has now less data from each speaker, it was not able to exploit these
adapted features properly.

The proposed model, on the other hand, outperforms both baselines in 3 out of 4 metrics, only
achieving similar scores in the SRMR metric. Compared to Wu2016, our model leads
to improvements of around 0.4 in PESQ, 0.1 in STOI, and 0.5 in POLQA.

% Comparison between results with single-speaker dataset (IEEE) vs.
% multi-speaker (either VCTK or TIMIT)

\subsection{Real vs. simulated RIR}
\begin{figure*}
    \centering
    \subfloat[(a)]{%
        \includegraphics[width=.49\textwidth]{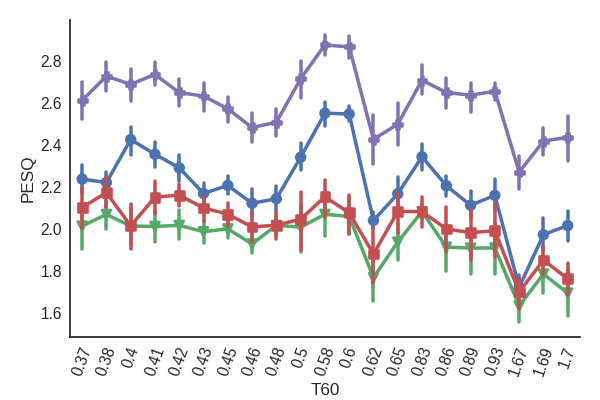}}
    \hfill
    \subfloat[(b)]{%
        \includegraphics[width=.49\textwidth]{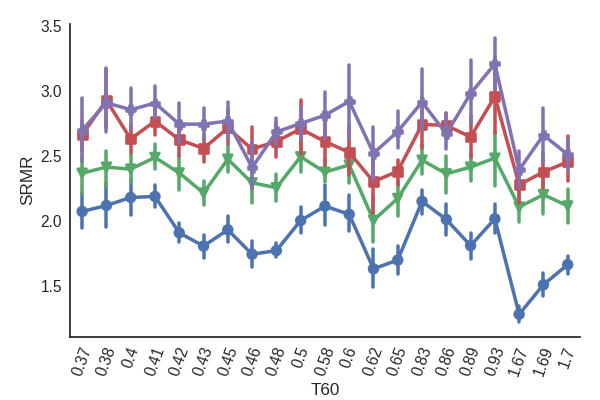}}
    \hfill
    \subfloat[(c)]{
        \includegraphics[width=.49\textwidth]{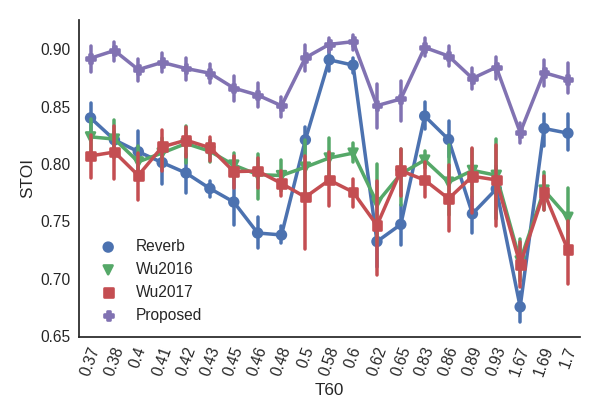}}
    \hfill
    \subfloat[(d)]{%
        \includegraphics[width=.49\textwidth]{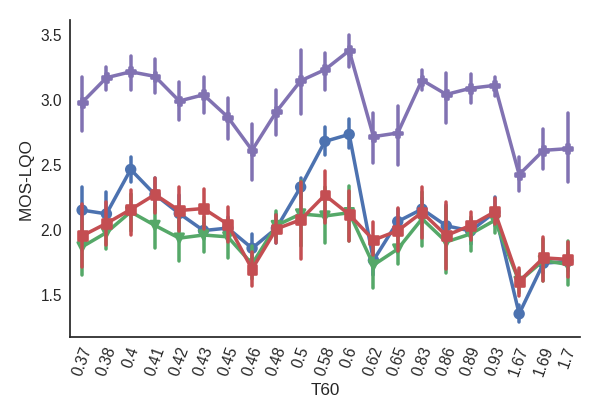}}
    \hfill
    \caption{Scores for experiment with real RIRs: (a) PESQ, (b) SRMR,
    (c) STOI, and (d) POLQA.}
    \label{fig:exp4}
\end{figure*}

In our last experiment, we test the generalization capability of the models
trained on simulated RIRs to real RIR. To that end, we used speech convolved
with real RIRs from the ACE Challenge dataset, as specified in Section IV. The
results are reported in Figure \ref{fig:exp4}. Note that the x-axis in that
figure does not have linear spacing in time, as we are showing the results for
each T60 as a single point uniformly spaced from its nearest neighbours in the
data. Note also the larger variability in scores for reverberant files, which
is due to the scores here not being averaged across many RIRs.

In this test, the proposed method achieves the highest scores in all metrics.
It is also the only method to improve PESQ and POLQA across all scenarios,
while the baselines either decrease or do not improve such metrics.  Although
the baselines do improve STOI in most cases, in some scenarios they actually
decrease STOI.

\subsection{Subjective Listening Tests} 
The results of MUSHRAR tests are summarized in Table \ref{tab:mushrar}. The results for the original, unprocessed reverberant files are also reported. When rating how reverberant the enhanced stimuli were, participants rated the outputs of the proposed model as less reverberant than the baseline and reverberant signals for all T60 values, while the baseline model only achieved a lower reverberation perception for lower reverberation times.

\begin{table}
\centering
\label{tab:mushrar}
\caption{Results of the MUSHRAR test for reverberation perception for different T60 values. Lower is better.}
\begin{tabular}{ccccc}
Model & 0.6~s & 0.9~s & 1.2~s & 1.5~s \\
\hline
Reverb & 42.62 & 60.06 & 71.73 & 82.24 \\
Wu2016 & 32.27 & 38.08 & 46.27 & 43.89 \\ 
Proposed & 24.44 & 23.89 & 29.93 & 31.64 \\
\end{tabular}
\end{table}

\section{Discussion}
\subsection{Effect of the context size and residual connections}
In this paper, we propose the addition of a few components to the architecture of deep neural networks for dereverberation. Using a context window for speech enhancement is not a new idea, and most studies using deep neural networks use similar context sizes. The novelty in our approach, however, is to use a convolutional layer as a local context encoder, which we believe helps the model to learn how to extract local features both in the time and the frequency axes in a more efficient way than just using a single context window as the input for a feedforward model. This local context, together with the residual input connections, is used as an input for the recurrent decoder, which is able to learn longer-term features. Similar architectures (save for the residual connections) have been successfully used for speech recognition tasks \cite{hannun_deepspeech_2014} but, to the best of our knowledge, this is the first work where such an architecture is used with the goal of estimating the clean speech magnitude spectrum.

Regarding the effect of the context size, our results agree with the intuition that including more past and future frames would help in predicting the current frame. However, we also show that there is a very small difference between using 7 vs. 11 frames as context. A window of 7 frames encompasses a total of 144 ms, vs. 208 ms for 11 frames. Both sizes are still much smaller than most of the reverberation times being used for our study, but they already start allowing the model to easily extract features related to amplitude modulations with lower frequencies, which are very important for speech intelligibility. We believe the combination of short- and long-term contexts, by having both the short-term context encoder and longer-term features through the recurrent layers, allows our architecture to benefit even from a shorter context window at the input. Lower modulation frequencies can still be captured through the recurrent layers, although we cannot explicitly control or assess how long these contexts are since they are learned implicitly through training and might be input-dependant.

We also introduce the use of residual and skip connections in the context of speech enhancement. Residual connections have a very close corresponding method in speech enhancement, namely spectral subtraction. The output of a speech enhancement model is likely to be very similar to the input, save for a signal that has to be subtracted from it (e.g. in the case of additive noise). In our case, our model is not restricted to subtraction due to how our architecture was designed. Consider the architecture as shown in Figure 1. The input to each recurrent layer is the sum of projections of the corrupted input (via the linear layers $f_1$, $f_2$, and $f_3$), and the output of the previous layers (via the linear layers $g_1$, $g_2$, and $g_3$). One possible solution to the problem, given this architecture, is for each recurrent layer to perform a new stage of spectral estimation given the difference between what the previous stage has predicted and the input at a given time. The output layer uses skip connections to the output of each recurrent layer and combines their predictions, allowing each layer to specialize on removing different types of distortions or to successively improve the signal. Our architecture was inspired by the work on generative models by Alex Graves \cite{graves_generating_2013}, which uses a similar scheme of connections between recurrent layers. The recently proposed WaveNet architecture \cite{oord_wavenet:_2016}, which is a generative model for audio signals able to synthesize high-quality speech, also makes use of residual connections and skip connections from each intermediate layer and the output layer.

\subsection{Comparison with baselines}
As seen in the previous section, our model outperforms both baselines based on feed-forward neural networks. One clear advantage of our approach is that we do not need to predict T60, which is a hard problem in itself and adds another layer of complexity to the model, especially under the presence of other distortions such as background noise \cite{AceChallenge}. 

Regarding model size, as seen in Table 1, it is important to note that despite having a significantly smaller number of parameters than the baseline models, our proposed model consistently outperforms them in most experiments. Our best model (shown in the last line of Table 1) has approximately half the number of parameters but uses these parameters more efficiently because of its architectural characteristics.

Although the authors of \cite{wu_reverberation-time-aware_2017} argue that their model leads to improvements in PESQ scores, the differences reported were not significant. Also, the authors have used PESQ
scores for selecting the best models; however, that might be an issue,
especially because PESQ is not a recommended metric for reverberant/dereverberated
speech. In our experiments we used the best validation
loss (MSE) for model selection for all the models, including the baselines.
Using a proper speech quality or intelligibility metric as a function for model selection would be a good, but costlier solution because one has to generate/evaluate samples using that metric for all epochs, while MSE can be computed directly from the output of the model.
\subsection{Generalization capabilities}
Another important aspect of our study, when compared to \cite{wu_reverberation-time-aware_2017}, is that we train and test the models under several different room impulse responses, both real and simulated. The results we report for Wu2017 have much lower performance than those reported in \cite{wu_reverberation-time-aware_2017} for similar T60 values; however, it must be noted we used random room geometries and random source and
microphone locations for our RIRs, while in their experiments the authors used
a fixed room geometry with different T60 values (corresponding to different
absorption coefficients in the surfaces of the room) and a fixed location.
Although they tried to show the model generalizes to different room sizes, they
tested generalization by means of two tests: (a) a single different room size
with the same source-microphone positioning and (b) and a single different
source-microphone positioning with fixed room geometry. We believe those
experiments were not sufficient to show the generalization capabilities of the
model, and experiments 1 and 3 in our work confirm that hypothesis. In our work, we tried to expose the model to several different room geometries and source-microphone positioning, since this is closer to real-world conditions and helps the model to better generalize to unseen rooms and setups.

\subsection{Anedoctal comparison with mask-based methods}
Although we did not compare our method to masking alternatives in this study, we would like to briefly mention the results reported in the most recent paper with a masking approach \cite{williamson_time-frequency_2017}, which used a similar single-speaker dataset (IEEE sentences uttered by a male speaker), simulated and real RIR for the dereverberation task. Although the range of T60s in our study and theirs is not similar, we can roughly compare the metrics of our model in the same range used in their study. The STOI improvements for our method are higher than the so-called cRM method proposed in that paper: our method has a STOI improvement of approximately 0.2 in all simulated T60s they report (0.3, 0.6, and 0.9~s), while their highest improvement is of 0.06 for 0.9~s. The cRM method, however, slightly decreased STOI for a T60 of 0.3~s.

\subsection{Study limitations}
Although we report both PESQ and SRMR scores in this work, these results should
be taken cautiously. PESQ has been shown to not correlate well with reverberant
and dereverberated speech. SRMR, on the other hand, is a non-intrusive metric,
so it is affected by speech- and speaker-related variability \cite{Santos_NH_2014}. It is also more sensitive to high reverberation times (0.8~s as shown in \cite{senoussaoui_speech_2017}), so measurements in low reverberation times
might not be accurate enough for drawing conclusions about the performance of a given model.

A characteristic common to all DNN-based models and also other algorithms based on spectral magnitude estimation is a higher level of distortion due to reusing the reverberant phase, especially in higher T60. This is added to magnitude estimation errors which are also higher in higher T60, since the input signal is very different from the target due to the combined effect of coloration and longer decay times. Although we do not try to tackle this issue here, there are recent developments in the field that could be applied jointly with our model, such as the method recently proposed in \cite{mayer_impact_2017}. This exploration is left for future work.

\section{Conclusion}
We proposed a novel deep neural network architecture for performing speech
dereverberation through magnitude spectrum estimation. We showed that this
architecture outperforms current state-of-the-art architectures and generalizes
over different room geometries and T60s (including real RIR), as well as to
different speakers. Our architecture extracts features both in a local context (i.e.,
a few frames to the past/future of the frame being estimated) as well as long-term
context. As future work, we intend to explore improved cost functions (e.g., incorporating sparsity in the outputs \cite{Jukic2017}) as well as applying the architecture to signals distorted with both additive noise and reverberation. We also intend to propose a multichannel extension of the architecture in the future. Finally, we intend to explore a number of solutions to the issue of reconstructing the signal using the reverberant phase.

\section*{Acknowledgement}
The authors would like to acknowledge funding from NSERC, FQRNT, and Google, as well as NVIDIA for the donation of the Tesla K40 GPUs used for the experiments.

\ifCLASSOPTIONcaptionsoff
  \newpage
\fi



\bibliographystyle{IEEEtran}
\bibliography{references.bib}

\begin{IEEEbiographynophoto}{João Felipe Santos}
received his bachelor degree in electrical engineering from the Federal University of
Santa Catarina (Brazil) in 2011 and an M.Sc. degree in telecommunications from the Institut National de
la Recherche Scientifique (INRS) in 2014, where he entered the Dean’s honour list and was awarded the
Best M.Sc. Thesis Award. He is currently a Ph.D. candidate in telecommunications at
the same institute. His main research interest is in applications of deep learning to speech and audio signal processing applications (speech enhancement, speech synthesis, speech recognition, and audio scene classification). 
\end{IEEEbiographynophoto}

\begin{IEEEbiographynophoto}{Tiago H. Falk}
(SM’14) received the B.Sc.
degree from the Federal University of Pernambuco, Brazil, in 2002, and the M.Sc. and Ph.D.
degrees from Queen’s University, Canada, in 2005
and 2008, respectively, all in electrical engineering. In 2010, he joined INRS, Montreal, Canada,
in 2010, where he is currently an Associate Professor and heads the Multimedia/Multimodal Signal Analysis and Enhancement Laboratory. His research interests include multimedia/biomedical
signal analysis and enhancement, pattern recognition, and their interplay in
the development of biologically inspired technologies.
\end{IEEEbiographynophoto}

\end{document}